# Probing Properties of Cold Radiofrequency Plasma with Polymer Probe

EDWARD BORMASHENKO[1,2]*, GILAD CHANIEL[1,3], VICTOR MULTANEN[1].

[1]Ariel University, Physics Faculty, 40700, P.O.B. 3, Ariel, Israel (edward@ariel.ac.il)
[2]Ariel University, Chemical Engineering and Biotechnology Faculty, 40700, P.O.B. 3, Ariel, Israel
[3]Bar Ilan University, Physics Faculty, 52900, Ramat Gan, Israel

**Abstract**

The probe intended for the characterization of cold plasma is introduced. The probe allows estimation of the Debye length of the cold plasma. The probe is based on the pronounced modification of surface properties (wettability) of polymer films by cold plasmas. The probe was tested with the cold radiofrequency inductive air plasma discharge. The Debye length and the concentration of charge carriers were estimated for various gas pressures. The reported results coincide reasonably with the corresponding values established by other methods. The probe makes possible measurement of characteristics of cold plasmas in closed chambers.

**Keywords:** cold plasma; sheath; concentration of charge carriers; polymer; probe; Debye length; contact angle.

1. **Introduction.**

Cold plasmas, produced by electrical discharges in low-pressure gases, consist of a mixture of highly reactive species, i.e. ions, radicals, electrons, photons and excited molecules (Lieberman and Lichtenberg, 2005). Their composition and characteristics strongly depend on device parameters, such as vacuum chamber geometry, gas pressure, gas flow rate and electrical power input and frequency (Lieberman and Lichtenberg, 2005). However, probing of physical properties of cold plasmas is a challenging task (Stenzel, 1976, Chasseriaux *et al*. 1972). When the cold plasma is open, such probing may be successfully carried out with the Langmuir probe (Hopwood, *et al*., 1993). When the cold plasma is enclosed in a chamber the probing procedure becomes far from to be trivial. At the same time, the precise data concerning the physical properties of plasma (especially the concentration of charged particles) is crucial for a diversity of applications of cold plasma, in particular, materials science application of plasmas.

Cold plasma treatment is broadly used for modification of surface properties of polymer materials (Yasuda, 1984; Strobel *et al*. 2000). The plasma treatment creates a complex mixture of surface functionalities, which influence surface physical and chemical properties and results in a dramatic change in the wetting behaviour of the surface (Occhiello *et al*. 1991, Occhiello *et al*. 1992, France and Short, 1997, France and Short, 1998, Wild and Kesmodel, 2001, Kondoh and Asano, 2000, Hegemann *et al*. 2003; Kaminska *et al*. 2002). Surface modification of polymers by cold plasmas allowed various applications of plasma for cleaning, printing and coating processes (Lieberman and Lichtenberg, 2005, Yasuda, 1984; Strobel et al. 2000).

In our paper we introduce the polymer-based probe, exploiting surface functionalization of polymers by the cold plasma, allowing effective characterization of plasma in closed chambers.

## 2. The role of a sheath in the interaction of cold plasma with polymer materials.

At the edge of a bounded plasma, a potential exists to contain the more mobile charged species. This allows the flow of positive and negative carriers to the wall to be balanced. In the usual situation of the plasma, consisting of equal numbers of positive ions and electrons, the electrons are far more mobile than the ions. The plasma will therefore be charged positively with the respect to a grounded wall. The non-neutral potential region between the plasma and the wall is called a sheath (Lieberman and Lichtenberg, 2005). Spatial distributions of charge carriers and a potential in the vicinity of a sheath for the DC plasma are depicted in Fig. 1.

When a polymer sheet (film) is placed into chamber containing a cold plasma, it is also surrounded by the sheath (polymer sheet works as a wall), as shown schematically in Fig. 2. The thickness of the sheath is on the order of magnitude of the few Debye lengths of the plasma given by:

$$\lambda_{De} = \left(\frac{\varepsilon_0 T_e}{e n_0}\right)^{1/2} \tag{1}$$

where $T_e$ is the temperature of electrons in units of Volts, $e$ is the charge of electron, $n_0 = n_i = n_e$ is the density of charge carriers far from a wall, depicted in Fig. 1 (Lieberman and Lichtenberg, 2005). The experimental establishment of $n_0$ in a closed chamber containing plasma is a challenging experimental task. We introduce the probe allowing the estimation of parameters of plasma based on surface effects produced by cold plasma on polymer films. We assume that the ion density is uniform and constant in time everywhere in the plasma. This is the good approximation for the cold radiofrequency plasma (Lieberman and Lichtenberg, 2005).

## 3. Probe allowing experimental estimation of parameters of cold plasma.

It is well known, that the cold plasma modifies strongly surface properties of organic materials, resulting in the dramatic change of their wettability (Strobel et al, 1994, Bormashenko and Grynyov, 2012). It is also generally accepted that the strong hydrophilization of polymer surfaces by cold plasma is at least partially due to the re-orientation of hydrophilic groups of polymers towards the solid/air interface, occurring under the cold plasma treatment (Yasuda and Sharma 1991, Mortazavi and Nosonovsky 2012, Bormashenko et al. 2013). This re-orientation results in the increase of the dipole-dipole interaction between the plasma-treated polymer and a liquid, contacting with this polymer (Kaminska et al. 2002, Pascual et al. 2008, Bormashenko et al. 2013). It is reasonable to attribute the re-orientation of hydrophilic groups of polymer chains to the collisions of these groups with ions accelerated by the electric field of a sheath, displayed in Fig. 1 (Lieberman and Lichtenberg, 2005). The electrons of the sheath practically do not transfer energy to much heavier hydrophilic moieties of polymer chains. Thus, it is plausible to suggest that when the sheath is not formed, and ions are not accelerated by the electric field of the sheath, the influence of plasma on a polymer will be essentially weakened.

This hints an idea of the plasma probe displayed in Fig. 3. This probe contains the glass slide and polymer film, placed into the aluminium frame, allowing free access of charged particles into the clearance separating the polymer and the glass slide. Now we place the probe into the chamber containing the cold plasma. When the clearance $h$ between the glass slide and polymer is larger than the Debye length $h > \lambda_{De}$, the sheath is formed and polymer will be influenced by ions of the sheath; however, when the condition $h < \lambda_{De}$ takes place the separation of charge carriers becomes impossible and a sheath will not be formed. Hence, when the condition $h < \lambda_{De}$ takes place the polymer will not feel the presence of plasma, and its surface properties will remain the same. Thus, varying the value of the clearance separating the polymer and the glass slide will allow the rough experimental estimation of the the Debye length $\lambda_{De}$. Thus, the concentration of charge carriers may be calculated with Exp. (1).

This hypothesis was experimentally checked with two polymers: polypropylene (PP) and polyethylene (PE). Extruded PP (thickness 30 *μm*) and PE (thickness 500 *μm*) films were used in our experiments. The linear dimensions of both of polymer films were the same, i.e. length – 5 *cm*, width – 2.5 *cm*. The probe was introduced into the inductive radiofrequency cold plasma discharge (Harrick PDC-32G). We performed experiments two series of experiments with various pressures of the air plasma *1* and 2*Torr*, the power in both of series was 18*W*. The time span of irradiation was 1min. The clearance *h* was varied from 10*μm* to 1400 *μm*. The simplest possible experimental procedure allowing detecting of plasma impact on polymer films is the measurement of the apparent water contact angle (Kaminska *et al*. 2002, Pascual *et al*. 2008, Bormashenko *et al*. 2013). The measurement of water contact angle was carried out under ambient conditions with a Ramé–Hart goniometer (model 500), immediately after introducing the probe into the chamber. Drops were placed at the center of polymer plates exposed to plasma. Fifteen measurements were taken to calculate the mean apparent water contact angle.

The apparent "as placed" (Tadmor and Yadav, 2008) water contact angle *θ* was plotted as a function of the clearance *h*, as shown in Fig. 4. We start the discussion from the results, obtained under the pressure of *2 Torr*. When the clearance *h* was larger than 150 *μm* the apparent contact angle of polymer films decreased dramatically. Starting from the critical value of $h_{cr}$~100-120 *μm* the impact of plasma on polymers was weakened, and when the clearance was less than 50 *μm*, no influence of plasma on wetting properties of polymers was registered, as demonstrated in Fig. 4. It should be emphasized that the results were the same for both of polymers used in our investigation, as it is seen from Fig. 4. This means that the critical value of clearance $h_{cr}$ depends on the properties of the plasma sheath only. Thus, the thickness of the sheath may be estimated roughly as $d \sim h_{cr} \sim 100$ *μm*. Thus, for the rough estimation of the concentration of charge carriers far from a wall $n_0$, we obtain using Exp. (1):

$$n_0 \approx \frac{\varepsilon_0 T_e}{e \lambda_{De}^2} \approx \frac{4 \varepsilon_0 T_e}{e d^2} \cong 2 \times 10^{16} \div 2 \times 10^{17} m^{-3} \qquad (2)$$

where $\lambda_{de} \approx d/2 \approx h_{cr}/2 = 50\,\mu m; T_e \cong 1 \div 10V$ are taken for a sake of estimation of $n_0$. The concentration of charge carriers, supplied by Exp. (2), coincides reasonably with the values typical for inductive cold radiofrequency plasma discharges (Hopwood, J. 1992, Hopwood, et al., 1993). Within the series of experiments performed under the lower pressure of *1 Torr* the critical value of $h_{cr}$ was established as 1300 μm. Correspondingly, the Debye length was established as: $\lambda_{de} \approx h_{cr}/2 = 650\,\mu m$, and the concentration of charge carriers was estimated as: $n_0 \approx 10^{14} \div 10^{15}\,m^{-3}$. Actually it is well expectable that decreasing of the gas pressure results in the diminishment of the concentration of charge carriers (Lieberman and Lichtenberg, 2005).

## 4. Summary.

We introduced simple probe intended for the characterization of cold radiofrequency plasma. The probe is based on the change of surface properties of polymer films caused by the cold plasma. This change is mainly due to the formation of plasma sheath surrounding a polymer film. The sheath accelerates ions colliding with a surface of polymer film. These collisions result in a dramatic change of wettability (hydrophilization) of polymer films. The proposed probe contains the glass slide and a polymer film, placed into the aluminum frame, allowing a free access of charged particles into the clearance separating the polymer and the glass slide. When the clearance between the glass slide and polymer is smaller than the Debye length, the sheath is not formed within the clearance. Hence, in this case the wettability of polymer will remain unchanged. This hypothesis was tested experimentally with use of polyethylene and polypropylene films. Indeed, starting from the certain threshold, minimal value of a clearance, the apparent water contact angle of polymer remained unchanged. The thickness of the sheath is on the order of magnitude of few Debye lengths. Thus the Debye length of the plasma was estimated. In our experiments carried out with inductive radiofrequency cold plasma the Debye length was estimated approximately as *50 μm* for the pressure of *2 Torr*, and *650 μm* for the pressure of *1 Torr*. The concentration of charge carriers was established respectively as: $n_0 \cong 2 \times 10^{16} \div 2 \times 10^{17}\,m^{-3}$ for the pressure of *2 Torr*, and $n_0 \approx 10^{14} \div 10^{15}\,m^{-3}$ for the pressure of *1 Torr*. This value coincides reasonably with the values of concentrations typical for cold radiofrequency plasma inductive discharges, established by other methods. It is noteworthy that the probe allows measurement of characteristics of cold plasma in closed chambers.

**Acknowledgements**

We are thankful to Mrs. Y. Bormashenko and Mrs. Al. Musin for their kind help in preparing this manuscript.

REFERENCES

H. K. Yasuda, J. Wiley & Sons 1984 Plasma polymerization and plasma treatment, New York.

Yasuda, H.; Sharma, A. K. 1981 Effect of orientation and mobility of polymer molecules at surfaces on contact angle and its hysteresis. *J. Polymer Sci*. **19**, 1285-1291.

M. Strobel, C. S. Lyons, K. L. Mittal. 1994 Plasma surface modification of polymers: Relevance to adhesion. VSP, Utrecht.

M. Strobel, C. S. Lyons, K. L. Mittal (Eds), 2000 Plasma surface modification of polymers: relevance to Adhesion. **Vol. 2**, VSP, Zeist, the Netherlands.

Stenzel, R. L. 1976 Microwave resonator probe for localized density measurements in weakly magnetized plasmas, *Rev. Scientific Instruments*. **47**, 603-607

Chasseriaux, J.M., Debrie, R, and C. Renard C. 1972, Electron density and temperature measurements in the lower ionosphere as deduced from the warm plasma theory of the h.f. quadrupole probe, *J. Plasma Physics.* **8**, 231 - 253

Lieberman, M. A.; Lichtenberg A. J. 2005 Principles of plasma discharges and materials processing, *J. Wiley & Sons*, Hoboken, (2005).

Hopwood, J. 1992 Review of inductively coupled plasmas for plasma processing, *Plasma Sources Sci. Technol.* **1**, 109.

Hopwood, J. Guarnieri, C.R. ; Whitehair, S.J. ; Cuomo, J.J. 1993, Langmuir probe measurements of a radio frequency induction plasma, *J. Vacuum Science & Technology A.* **11**, 152 – 156

Occhiello, M. Morra, F. Garbassi,1991 *Applied Surface Science*. **47**, 235-242.

Occhiello, M. Morra, P. Cinquina, F. Garbassi, 1992 *Polymer*. **33**, 3007-3015.

R. M. France, R. D. Short, 1998 *Langmuir*. **14 (17)**, 4827–4835.

R. M. France, R. D. Short, 1997 *J. Chem. Soc., Faraday Trans*. **93**, 3173-3178.

S. Wild, L. L. Kesmodel, 2001  *J. Vac. Sci. Technology*. **19**, 856-860.

Kondoh, T. Asano, A. Nakashima M. Komatu, 2000  *J. Vac. Sci. Technology*. **18**, 1276-1280.

D. Hegemann, H. Brunner, Ch. Oehr, 2003 Nuclear Instruments and Methods in Physics Research B **208**, 281–286.

A. Kaminska, H. Kaczmarek, J. Kowalonek, 2002 *European Polymer Journal*. **38**, 1915–1919.

Ed. Bormashenko, R. Grynyov, 2012 Plasma treatment allows water suspending of the natural hydrophobic powder (lycopodium), *Colloids and Surfaces B*. **97**, 171– 174

Ed. Bormashenko, G. Chaniel, R.Grynyov, 2013 Towards understanding hydrophobic recovery of plasma treated polymers: Storing in high polarity liquids suppresses hydrophobic recovery, *Applied Surface Science*. **273**, 549– 553

M. Pascual, R. Balart, L. Sanchez, O. Fenollar, O. Calvo, 2008 *J. Materials Science*. **43**, 4901 4909.

R. Tadmor, Pr. S. Yadav, 2008 As-placed contact angles for sessile drops, *J. Colloid and Interface Science*. **317**, 241–246.

M. Mortazavi, M. Nosonovsky, 2012 A model for diffusion-driven hydrophobic recovery in plasma treated polymers, *Applied Surface Science*. **258**, 6876–6883

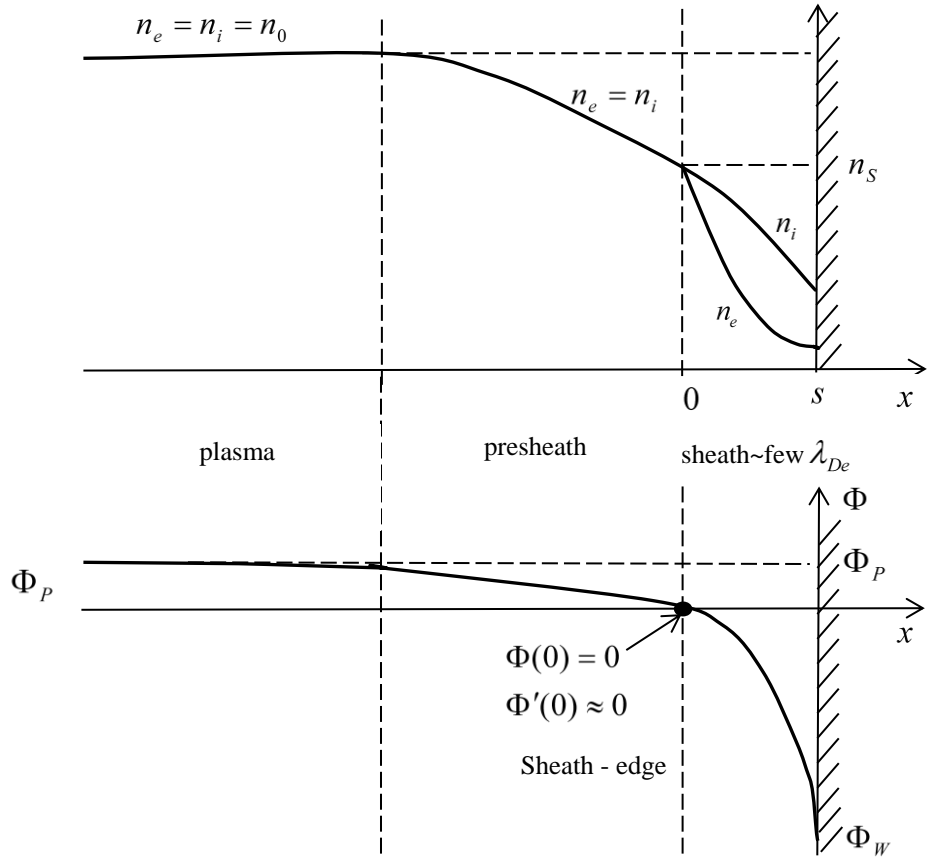

FIGURE 1. Scheme of a sheath in a contact with the wall for the DC plasma; $n_e; n_i$ are the concentrations of electrons and ions respectively, $\Phi$ is the potential; $\Phi_P; \Phi_W$ are the potentials of the plasma and wall respectively; $\lambda_{De}$ is the Debye length.

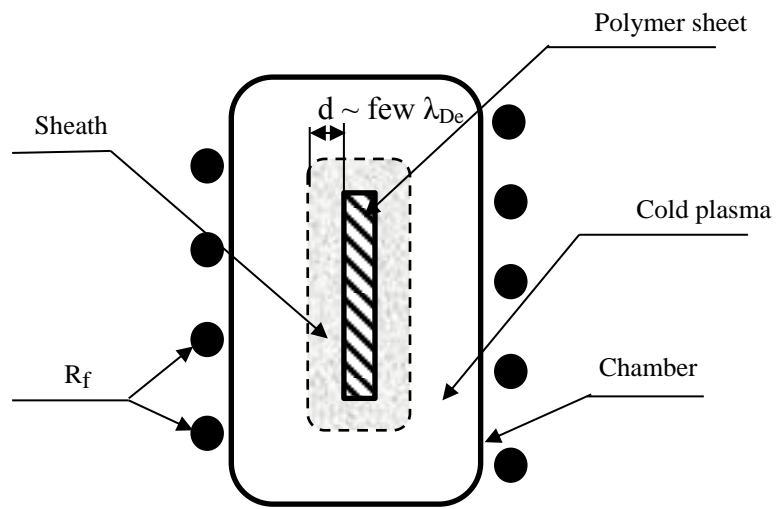

FIGURE 2. Polymer sheet introduced in a chamber containing cold plasma is surrounded by a sheath with thickness of *d*.

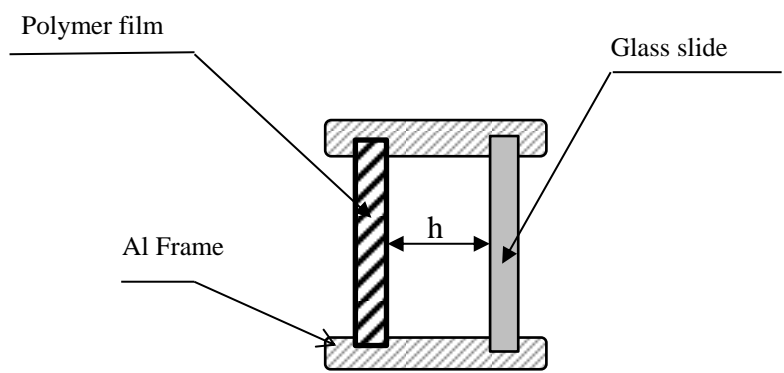

FIGURE 3. A sketch of the probe; the dimension of polymer films were 2.5x5 *cm*.

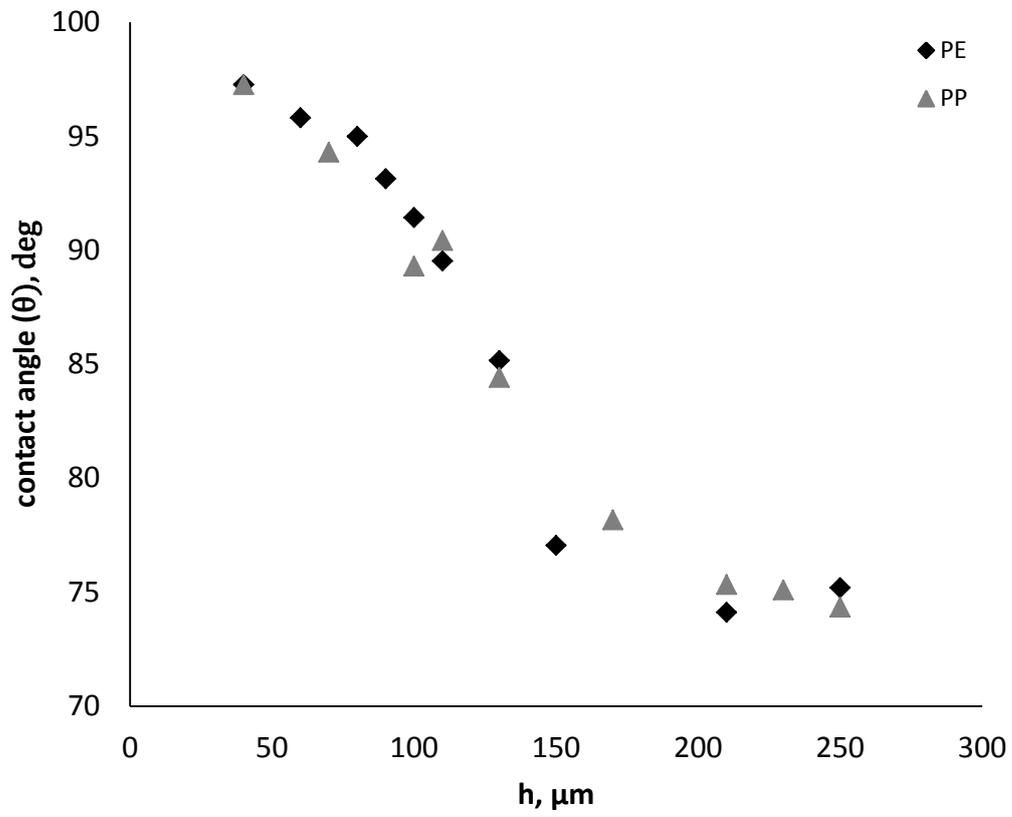

FIGURE 4. The apparent contact angle as function of the clearance *h*. The air pressure - *2 Torr*.